\documentclass[]{aa}
\usepackage{graphicx}
\usepackage{txfonts}
\usepackage{natbib}
\bibpunct{(}{)}{;}{a}{}{,}

\begin{document}

\title{Acceleration and radiation of ultra-high energy protons \\ in galaxy clusters}

\author{G. Vannoni\inst{1}\thanks{work performed as IMPRS fellow at the Max-Planck-Institut f\"ur Kernphysik, Heidelberg - \email{giulia.vannoni@cea.fr}} \and F. A. Aharonian\inst{2,3} \and  S. Gabici\inst{2} \and S. R. Kelner\inst{3,4} \and A. Prosekin\inst{3,4}}

\institute{ Commissariat \`a l'\'Energie Atomique, DSM / Irfu / SPP, 91191 Gif-sur-Yvette, France;
\and Dublin Institute for Advanced Studies, 31 Fitzwilliam Place, Dublin 2, Ireland;
\and Max-Planck-Institut f\"ur Kernphysik, Saupfercheckweg 1, Heidelberg 69117, Germany;
\and Moskow Institute of Engineering Physics, Karshiskoe sh. 31, Moskow, 115409, Russia.}

\abstract
{Clusters of galaxies are believed to be capable to accelerate  protons  at accretion shocks
to  energies exceeding   $10^{18}$ eV. At these energies, the losses caused 
by  interactions of cosmic rays  with  photons of the  Cosmic Microwave Background
Radiation (CMBR)  become effective and determine the maximum energy of protons 
and the shape of the energy spectrum  in the cutoff region.}
{The aim of this work is the  study of the formation of the energy spectrum of 
accelerated protons at  accretion shocks of galaxy clusters and of the 
characteristics of  their broad band emission.} 
{The proton  energy distribution is calculated  self-consistently 
via a time-dependent numerical treatment of the shock acceleration process 
which takes into account the proton energy losses due to interactions with the CMBR.}
{We calculate the energy distribution of accelerated protons, as well as the 
flux  of broad-band emission produced  by secondary electrons and positrons via synchrotron and inverse Compton scattering processes. We find that the downstream and upstream regions contribute almost at the same level to the emission. For the typical parameters characterising  
galaxy clusters, the synchrotron and IC peaks in the  spectral energy distributions appear  at comparable flux levels.}
{For an efficient acceleration, the expected emission components in  the X-ray and gamma-ray band  are close to the detection threshold 
of  current  generation instruments, and will be possibly detected with the future generation of detectors.}

%\authorrunning
%\titlerunning
\maketitle

\section{Introduction}
\label{Introduction}

Rich clusters of galaxies are the largest virialised structures in the Universe, with typical sizes of a few Mpc and masses up to $10^{15} M_{\odot}$ or more (see \citealp{sarazinbook} for a review).
In the standard picture of cosmic structure formation, the structure's growth is driven by gravitational instability. This process is hierarchical, with larger systems forming later via the assembly of pre-existing smaller structures.
Within this scenario, galaxy clusters form via mergers, and their age can be estimated to be of the order of 10 Gyr (e.g. \citealp{agecl}).
In addition, cold material from the surrounding environment is continuously infalling, due to gravitational attraction, and an expanding shock wave, called the accretion shock, is expected to form at the cluster boundary and to carry outward the information of virialisation \citep{bertschinger}. Numerical simulations have confirmed the appearance of the so--called accretion shocks during structure formation (e.g. \citealp{kangsim}).

The detection of a tenuous and diffuse synchrotron radio emission from about one third of rich clusters of galaxies (e.g. \citealp{govoni}) reveals the presence of a diffuse magnetic field and a population of high energy electrons. Using exclusively synchrotron data, though, it is possible to get information only on the product of particles density and magnetic field energy density (unless adopting assumptions like equipartition). Detection of non-thermal X-rays have also been claimed from a few of such sources and generally interpreted as inverse Compton emission from the same population of electrons (e.g. \citealp{ff, eckert}). Since IC depends on the electron density but not on the magnetic field, combining the data allows us to break the degeneracy and thus determine values of B of the order of $0.1 \mu$G. Significantly higher values, of the order of a few microGauss, are obtained from Faraday rotation measurements \citep{Clarke,carilli}. Thus the uncertainty in the determination of the magnetic field is quite large, allowing values to vary of about one order of magnitude \citep{nnrephaeli}.

According to theoretical models, the electrons responsible for the radio and non-thermal X-ray emission, are produced through different acceleration mechanisms (e.g. \citealp{schlickeiser,tribble,sarazin,brunetti2001,petrosian2001}). Alternatively, the radio synchrotron radiation  can be emitted by secondary electrons produced at interactions of accelerated protons with the intracluster gas \citep{dennison,blasicol,atoyan,dolag,kushnir}. In this scenario, gamma ray emission is expected due to the decay of neutral pions produced in \textit{p-p} interactions \citep{volk_cluster,ber_cluster}. The non-thermal X-rays emission also can be related to synchrotron radiation of secondary electrons, but of much higher energies, produced in photon-photon  \citep{timokhin} and proton-photon \citep{FA02,susumu} interactions.

Several particle acceleration mechanisms have been proposed to operate in clusters of galaxies (see \citealp{revste} for a review). In particular, it has been argued that large scale shocks can effectively accelerate electrons and protons up to ultrarelativistic energies \citep{norman,ber_cluster,loeb,miniati,blasimerger,steshocks,ryu,berrington,pfrommer}.
In a recent paper we performed  detailed calculations  of diffusive shock acceleration of electrons in galaxy clusters \citep{mio}. Electrons can be accelerated up to 100 TeV at cluster accretion shocks,  with synchrotron X-ray and  IC gamma-ray radiation components produced mainly in the downstream region. While the maximum energy of electrons is limited by synchrotron and IC energy losses, the protons can be accelerated to much higher energies.

Indeed, according to the Hillas criterion \citep{hillas},  galaxy clusters are  amongst the few source populations
capable, as long as this concerns the dimensions of the structure and the value of the magnetic field, to accelerate protons up to $10^{20}$ eV.  Moreover, clusters are cosmological structures and their lifetime is comparable to the age of the Universe, therefore, if acceleration takes place in such objects, it can continue up to $\sim 10^{10}$ yr.
On the other hand, high energy protons up to energies of the order of $10^{15}$ eV and possibly higher are well confined in the volume of the cluster over this time scale \citep{volk_cluster,ber_cluster}.
This results in an effective accumulation of high energy particles in the cluster. The thermal energy budget of rich clusters, estimated from measurements of their thermal X-ray emission, is of the order of $10^{63}-10^{64}$ erg. The non-thermal component seems to be constrained to a few percents of the thermal energy $\sim 10^{62}$ erg \citep{veritas,coma,magic}. This value is also comparable with the magnetic field energy, assuming 1 $\mu$G and a spherical cluster of 3 Mpc of radius (e.g. the size of the Coma cluster).

Although the dimensions of the system, the strength of the magnetic field,  and the age of the accelerator formally allow protons to be accelerated up to $10^{20}$ eV, the particles lose, in fact,  their 
energy via pair and pion production in the interactions with the photons of the Cosmic Microwave Background (CMBR) radiation field. As discussed in \citet{norman} and \citet{kang_rach_bier}, for a shock velocity of a few thousands km/s and a magnetic field of the order of  $1 \mu \rm G$, the shock acceleration rate is compensated by the energy loss rate at energies around $10^{19}$ eV (the exact value depends on the assumed diffusion coefficient and also on the shock geometry as shown by \citealp{ostrowski}).  The interactions of ultra-high energy protons with the CMBR lead to production of electrons in the
energy domain ($\geq 10^{15} \ \rm eV$) which is not  accessible through any direct acceleration mechanism. 
These electrons   cool  via synchrotron radiation and IC scattering   on very short  timescales 
(compared to  both the age of the source and  interaction timescales of protons).  For the same reason
they are rather  localised in space, i.e. are ``burned" not far from the sites of their production. 
Therefore the corresponding radiation components in the X-ray and gamma-ray energy bands 
are  precise tracers of primary protons, containing information about the acceleration and propagation 
of their ``grandparents". This interesting  feature has been indicated in  \citet{FA02} in the
general context of  acceleration and propagation of ultrahigh energy protons in large scale extragalactic structures. More specifically, this issue was discussed by \citet{susumu} for objects like the Coma galaxy cluster. However in these  studies  an {\it a priori} spectrum of protons has been assumed in the 
``standard form"  $E^{-2} \exp(-E/E^*)$. In fact,  the  energy distribution of protons accelerated 
in galaxy clusters via Diffusive Shock Acceleration (DSA) mechanism  can be quite different from this 
simple form, as it is demonstrated below. 

In this paper we  study the process of proton acceleration by accretion shocks in  
galaxy clusters taking into account self-consistently the  energy loss channels of 
protons related to their interactions with  the CMBR photons. 
We make use of the numerical approach presented in \citet{mio}. 
In Section \ref{Proton Acceleration and Energy losses} the calculation is introduced 
and the accelerated proton spectra are derived. In particular, we show that electron/positron 
pair production is the dominant energy loss channel for protons. In Section 
\ref{Spectra of Secondary Electrons} we calculate the spectra of  secondary pairs. 
The broad-band emission produced by the secondary electrons during their interactions with the background magnetic and radiation fields is presented in Section \ref{Radiation Spectra}. In Section \ref{Shock Modification} we study, in a simplified fashion, the impact of shock modification by efficiently accelerated protons on the acceleration and emission features.  We  briefly discuss and summarise the main results  in Section \ref{Conclusions}. 

\section{Proton Acceleration and Energy losses}
\label{Proton Acceleration and Energy losses}

The accretion of cold external material onto a hot rich cluster of galaxies can lead to the formation of a strong shock at the position of the virial radius of the cluster. The shock velocity can be estimated via the free fall velocity of the infalling matter crossing the shock surface, namely:
\begin{equation}
v_{s} \sim \sqrt{\frac{2 G M_{cl}}{R_{cl}}} \approx 2000 \left( \frac{M_{cl}}{10^{15} M_{\odot}} \right)^{1/2}  \left( \frac{R_{cl}}{3 Mpc} \right)^{-1/2}~\textrm{km/s}.
\label{velocity}
\end{equation}
It is interesting to note that the shock velocities for this case 
are comparable to the shock speeds typical for  young supernova remnants.

We approximate the spherical shock locally as plane and work in the 
reference frame where the shock is at rest and the plasma moves along the $x$-axis, perpendicular to the shock surface, from $- \infty$ far upstream, to $+ \infty$ far downstream and the shock is located at $x=0$. In the following, all the quantities in the upstream region will be indicated with the subscript 1, and all the quantities downstream with the subscript 2.
To obtain the accelerated proton spectrum under the effect of energy losses induced by the interaction with the CMBR, we adopt the numerical scheme described in \cite{mio}. The time dependent transport equation for the particle distribution function in phase space is:
$$\frac{\partial f(x,p,t)}{\partial t}+u\frac{\partial f(x,p,t)}{\partial x}-\frac{\partial }{\partial x}\left( D(x,p)\frac{\partial f(x,p,t)}{\partial x}\right)-\frac{p}{3}\frac{\partial u}{\partial x}\frac{\partial f(x,p,t)}{\partial p}$$
\begin{equation}
-\frac{1}{p^2}\frac{\partial }{\partial p}(p^2 L(x,p) f(x,p,t))=Q(x,p),
\label{trans}
\end{equation}
where $L(x,p)=-\dot p$ is the  energy  loss rate  and $Q(x,p)$ is the injection term; $u$ represents the bulk velocity of the plasma in the shock rest-frame. The momentum and space dependence of the energy loss term as well as of the diffusion coefficient can be chosen of any form so that the formulation is general. Eq. (\ref{trans}) is solved numerically, inserting the boundary conditions at the shock location and at upstream/downstream infinity. For the boundary at the shock we consider again Eq. (\ref{trans}) and integrate it between $x = 0_{-}$ immediately upstream and $x = 0_{+}$ immediately downstream,  leading to:

\begin{equation}
\frac{1}{3}(u_1-u_2)p\frac{\partial f_0}{\partial p}=D_2 \frac{\partial f_0}{\partial x}\Big|_2 - D_1 \frac{\partial f_0}{\partial x}\Big|_1+Q_0 \delta(p-p_0).
\label{boundary}
\end{equation}
At $\pm \infty$ we set $f(x,p)=0$.

We assume that injection happens at the shock surface and that it can be described by a delta-function in momentum:
$$Q(x,p)=Q_0\delta(x) \delta(p-p_0),$$
where $p_0$ is the injection momentum and $Q_0$ a normalisation constant.

%
%%%%%%%%%%%%%%%%% fig.1
\begin{figure}[]
\centering
\includegraphics[scale=0.4]{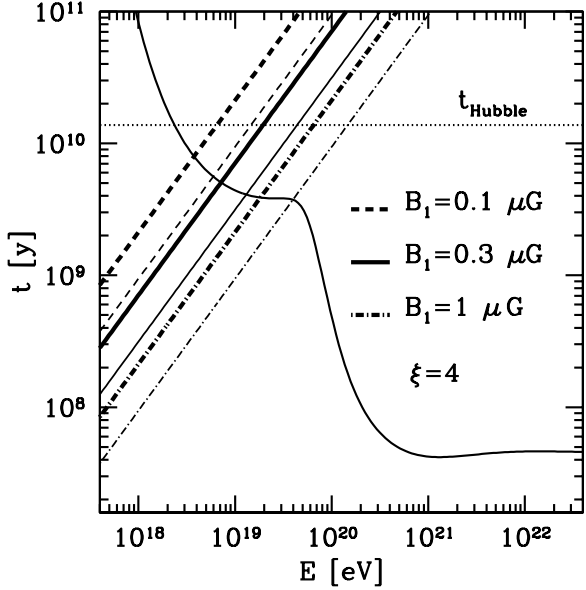}
\caption{\footnotesize{Acceleration and energy loss time scales as a function of the proton energy. The acceleration time scales are obtained for the values of the upstream magnetic field $B_1$ reported in figure and a downstream magnetic field $B_2=4B_1$. The thick lines correspond to a shock velocity of 2000 km/s, the thin lines to a velocity of 3000 km/s. As an horizontal dotted line we report the estimated age of the Universe, for comparison.}}
\label{fig:losPr}
\end{figure}
As pointed out in \cite{norman} and \cite{kang_rach_bier}, the relevant energy loss channel for the protons accelerated at cluster accretion shocks is the interaction with the CMBR.  Other radiation fields do not play any significant role in this regard. The interaction proceeds via two main processes of electron/positron pair production, also known as Bethe-Heitler  pair production, and photomeson production.  In Fig. \ref{fig:losPr}  we show the cooling  time-scales of protons 
due to these two processes  (solid curve).  Pair production dominates up to energies of about $5\times 10^{19}$ eV. Once the particle energy has passed the threshold for meson production, this latter  becomes rapidly dominant.  The thick lines in Fig. \ref{fig:losPr} correspond to the acceleration times for a shock velocity of 2000 km/s and an upstream magnetic field of 0.1 $\mu$G (dotted), 0.3 $\mu$G (dashed) and 1 $\mu$G (dash-dotted). For the same magnetic field, but for the shock speed of 3000 km/s,  the acceleration times are shown with  thin lines.  The downstream magnetic field is assumed of the form $B_2=\xi B_1$, where $\xi$ is the compression factor at the shock.  For a strong linear shock,  the parameter $\xi$  varies  between  1 to 4, depending on the orientation of the upstream magnetic field. We choose $\xi=4$, appropriate for Alfv\`enic turbulence. Fig. \ref{fig:losPr} shows that if protons can be accelerated beyond few times $10^{18}$ eV, the cut-off energy as well as the shape of the particle spectrum in the cut-off region are determined by pair production.  Note that  even for extreme cluster parameters,  the protons do not achieve energies for which 
the energy losses would be dominated by photomeson production.  

In this work we choose the following values of  model parameters: a shock velocity of 2000 km/s and a magnetic field upstream $B_1=0.3~\mu$G, with $B_2=4B_1$ (thick solid line in Fig. \ref{fig:losPr}). We do not take into account the evolution of the CMBR with redshift. This simplification is justified by the fact that the acceleration rate and the energy loss rate become comparable on a time scale of about 5 Gyr. 
Therefore, for local clusters, this time scale places the start  of the acceleration process at an epoch where the redshift is $z \ll 1$ and we can neglect the evolution of the CMB radiation. 
Moreover, as discussed in \citet{kang}, the higher temperature and energy density of the CMBR at large redshifts inhibits the acceleration of protons to ultrahigh energies at early epoch.

%%%%%%%%%%%%%%%%%fig.2
\begin{figure}[]
\centering
\includegraphics[scale=0.4]{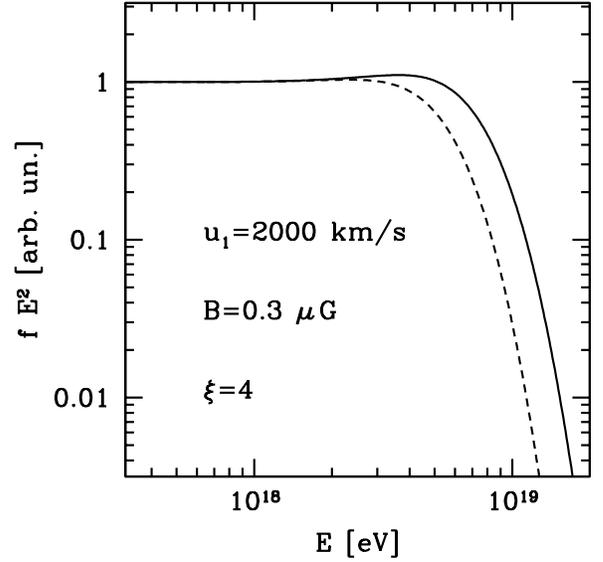}
\caption{\footnotesize{Proton spectra at the shock location for an acceleration time of 10 Gyr (solid line) and 5 Gyr (dashed) for a shock velocity of 2000 km/s, a magnetic field upstream $B_1=0.3$ $\mu$G and a magnetic field downstream $B_1=4B_1$.}}
\label{fig:shockPr}
\end{figure}

In Fig. \ref{fig:shockPr} we plot the calculated proton spectra at the shock location.
Our results are more conveniently shown as a function of the particle energy, rather than their momentum.
We perform our calculation for an age of the system of 5 Gyr and 10 Gyr. The difference between the spectra at the two ages demonstrates that the system does not reach steady state. Therefore, a time dependent calculation is essential in order to model correctly  the particle spectrum.
The cut-off energy is located around $7\times 10^{18}$ eV.
A small bump is present in the spectra around the cut-off energy, due to the flattening of the energy loss time scale in that energy range. The prominence of the feature increases with time.

%%%%%%%%%%%%%%%%%%% fig.3
\begin{figure}[]
\centering
\includegraphics[scale=0.4]{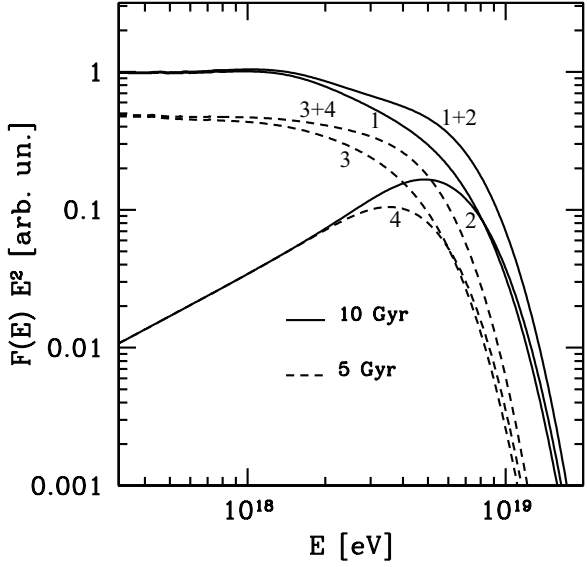}
\caption{\footnotesize{Spatially integrated spectra for the proton distributions in Fig. \ref{fig:shockPr} for an accelerator age of 10 Gyr (solid lines) and 5 Gyr (dashed lines). The lines represent: 1 and 3 the downstream contributions, 2 and 4 the upstream ones, and the sum of the two. }}
\label{fig:intPr}
\end{figure}
Once the particle distribution function is obtained for every point of the phase space, we proceed to calculate the spectrum integrated over space $F(E) = \int f(x,E)dx$, in order to obtain the production spectra of secondary pairs. The results are plotted in Fig. \ref{fig:intPr} for the up- and down-stream regions of the shock separately. The spectral features are evident in the spatially integrated spectra. At energies below the cut-off, the upstream particle spectrum is hard. This is due to the fact that particles can propagate ahead of the shock front over a distance of the order of their diffusion length, defined as $x_D=D(E)/u_1$. For Bohm-type diffusion $x_D \propto E$, and the integrated spectrum upstream is thus proportional to the particle spectrum at the shock multiplied by $E$. Downstream, on the other hand, at energies low enough that energy losses are negligible, the spectrum is $f_2(x,E)=f_0(E)\propto E^{-2}$. In this component, a bump forms around $10^{18}$ eV due to the fact that around that energy (close to the threshold of the process) the pair production loss rate changes behaviour from a very steep dependence on the particle energy to an almost flat distribution. Therefore, particles at slightly lower energies are only marginally affected by losses, as we can see from the fact that, below $10^{18}$ eV, a power law spectrum $\propto E^{-2}$ is recovered, as expected for an uncooled spectrum; on the other hand, particles at slightly higher energies experience a very fast cooling so that there is an effect of accumulation around the threshold. The effect is analogous to the one discussed in \citet{ber_dip1}, explaining the appearance of a bump in the ultrahigh energy Cosmic Ray spectrum due to photomeson production during propagation. Quite  interestingly, as a consequence of the shallow dependence on energy of the loss time scale at the cut-off energy and above it, the spectrum in the cut-off region is smoother than a simple exponential behaviour. At energies around $10^{20}$ eV, a second steepening is in fact present (not shown in figure), due to the effect of pion production, however at that energy the flux is suppressed by several orders of magnitude with respect to the cut-off value and thus meson production is negligible for changing the proton spectrum.

It is worth noting that, even though the contribution from the downstream region is dominant at low energies, around the cut-off the contribution of the two components differ only by a factor of few.

\section{Spectra of Secondary Electrons}
\label{Spectra of Secondary Electrons}

%%%%%%%%%%%%%%%%%%% fig.4
\begin{figure}[]
\centering
\includegraphics[scale=0.4]{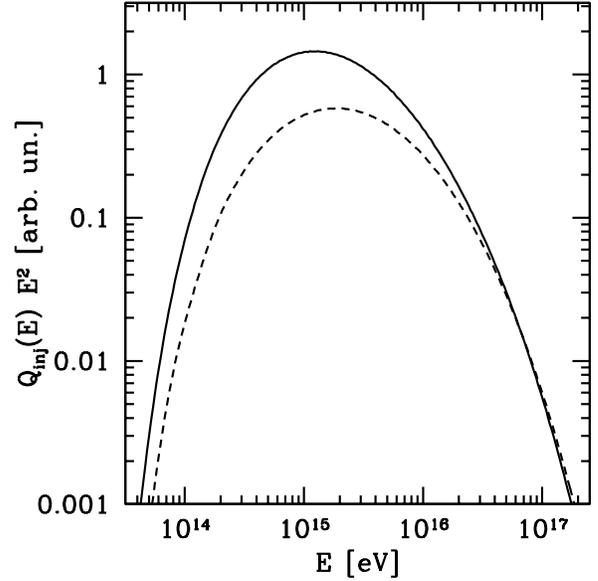}
\caption{\footnotesize{Production spectra of pairs. Solid line: downstream spectrum for an accelerator age of 5 Gyr; dashed line: upstream spectrum.}}
\label{fig:sptcoppie}
\end{figure}
In the following we present the results calculated for a cluster with an age of 5 Gyr. For simplicity we assume that the magnetic field is  homogeneously distributed over the cluster.   The formalism proposed 
by \citet{kelner} was used for calculations of the production spectra of  secondary  
electron-positron pairs.  Fig. \ref{fig:sptcoppie} shows the spectra of secondary electrons up and downstream produced by the proton spectra shown in Fig. \ref{fig:intPr}.

%%%%%%%%%%%%%%%%%%%%% fig.5
\begin{figure}[]
\centering
\includegraphics[scale=0.4]{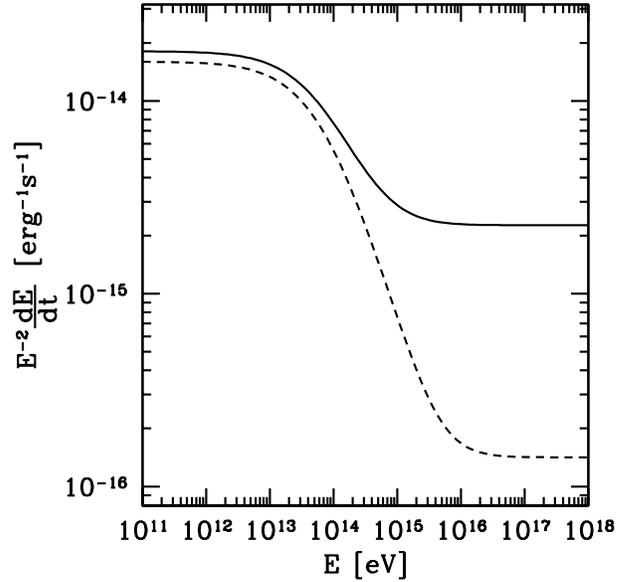}
\caption{\footnotesize{Energy loss rates for pairs due to synchrotron and IC processes. Solid line: downstream (magnetic field  $B_2=1.2~\mu$G), dashed line: upstream (magnetic field  $B_1=0.3~\mu$G).}}
\label{fig:losscoppie}
\end{figure}
Once produced, the electrons (hereafter the term ``electrons''  implies both electrons and positrons), lose energy  at interactions with the CMBR  and the ambient magnetic field. The energy loss rates for these two processes are shown  in Fig. \ref{fig:losscoppie}. The curves are multiplied by $E^{-2}$ in order to 
indicate the transition of the Compton losses from the Thomson to Klein-Nishina regime which takes place 
around 10 TeV. The curves correspond to the sum of synchrotron and IC energy loss rates in the downstream  and  upstream regions (solid and dashed line, respectively). In both cases the target photon field for IC is the CMBR with temperature 2.7 K. The magnetic field downstream is 4 times  larger than upstream.  While at low energies the Compton losses dominate over the synchrotron losses, at  very high energies, $E \geq 1000$~TeV,  because of the Klein-Nishina effect,  the  synchrotron losses become the dominant channel of  cooling of electrons.  

For the given injection (source) rate $Q_{inj}(E)$, the energy distribution of electrons $F(E)$ is 
described, in the continuous energy-loss approximation,  by the equation:
\begin{equation}
\frac{\partial F(E)}{\partial t}+\frac{\partial }{\partial E}\left(\frac{dE}{dt}F(E)\right)=Q_{inj}(E).
\end{equation}
Here we assume that the particles do not undergo any further acceleration during their lifetime.
In the steady state limit, $\partial F(E)/\partial t=0$,  the energy distribution of electrons 
is given by:
\begin{equation}
F(E)=\frac{1}{dE/dt}\int_E^{\infty} Q_{inj}(E') dE'.
\label{eq:cooledspt}
\end{equation}
The steady state solution correctly describes the spectrum of electrons, given that the lifetime of 
high energy  electrons is shorter than the age of the source.  

%%%%%%%%%%%%%%%%%%%% fig.6
\begin{figure}
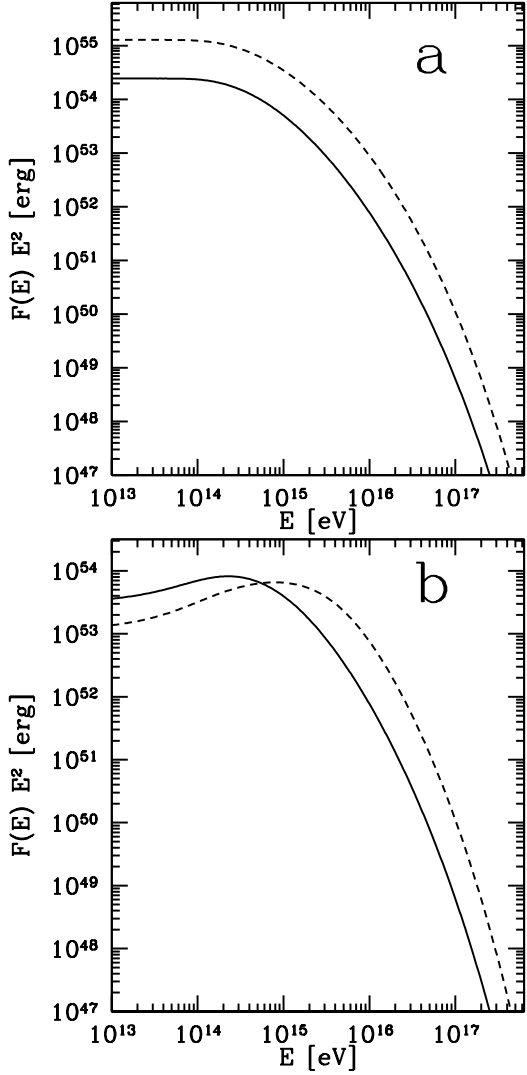

\begin{minipage}[t]{7.cm}
\centering
\includegraphics[scale=0.37]{./cooledSyn5GyrBWa.epsi}
\end{minipage}
\ \hspace{2mm} \hspace{3mm} \
\begin{minipage}[t]{7.cm}
\centering
\includegraphics[scale=0.37]{./cooledspt5GyrBWb.epsi}
\end{minipage}
\caption{\footnotesize{Electron spectra after taking into account energy losses. The solid and dashed lines refer to the downstream and upstream regions respectively and the spectra are normalised to a total energy in accelerated protons of $10^{62}$ erg in the volume of the cluster. \textbf{a}) Spectra obtained considering pure synchrotron cooling in a magnetic field $B_1=0.3~\mu$G upstream and $B_2=1.2~\mu$G downstream. \textbf{b}) Same as in panel \textbf{a} but with the realistic combination of IC and synchrotron losses as in Fig. \ref{fig:losscoppie}}}
\label{fig:cooledspt}
\end{figure}

The two panels in Fig. \ref{fig:cooledspt}  show the cooled spectra of electrons downstream (solid) and upstream (dashed). The normalisation is obtained assuming a total energy in accelerated protons of $10^{62}$ erg in the volume of the cluster.
The spectra are calculated under two assumptions:  pure synchrotron losses (panel \textbf{a}),
and synchrotron plus IC losses (panel \textbf{b}). The case in panel \textbf{a} is not realistic, but it 
allows us to understand the effects introduced by IC losses.  Because of very hard injection 
spectrum of electrons below 100  TeV (see  Fig. \ref{fig:sptcoppie}), the synchrotron  losses
lead to a standard $E^{-2}$ type  steady state spectrum of electrons with a high-energy cutoff which is determined by the cutoff in the 
injection spectrum.  Although  the production rates  of pairs upstream and downstream 
are quite similar,  more intense synchrotron losses downstream suppress the electron  
spectrum in that region. The effect of IC losses, which above 100 TeV takes place in the KN regime,  leads to a more complex feature, in particular to a hardening below the cut-off. 
Since  IC  losses affect the two sides of the shock in the same way, the slightly higher production rate of pairs downstream  results  in a  higher  flux of cooled electrons at energies up to $10^{15}$ eV . At higher energies  the synchrotron losses become dominant (see Fig. \ref{fig:losscoppie}), and thus  determine the spectral shape of electrons.

\section{Radiation Spectra}
\label{Radiation Spectra}

We can now calculate the emitted radiation spectra based on the electron distributions in Fig \ref{fig:cooledspt}\textbf{b}.
%%%%%%%%%%%%%%%%%%%% fig.7
\begin{figure}[]
\centering
\includegraphics[scale=0.5]{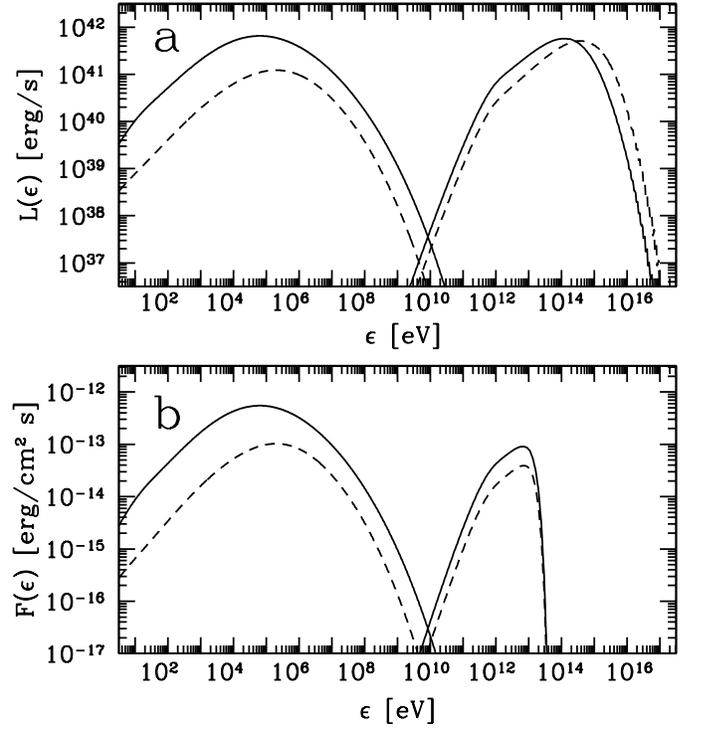}
\caption{\footnotesize{\textbf{a}) Broadband electromagnetic emission produced at the source location via synchrotron and IC cooling by the electron distributions in Fig. \ref{fig:cooledspt}\textbf{b}. The solid curve refers to the downstream component and the dashed curve to the upstream one. \textbf{b}) Expected flux at Earth for a source located at a distance $D=100$ Mpc. The effect of photon-photon absorption during propagation has been taken into account.}}
\label{fig:radcoppie}
\end{figure}
The broadband Spectral Energy Distribution (SED) of photons produced via synchrotron and IC emission is shown in Fig. \ref{fig:radcoppie}\textbf{a}. Solid and dashed lines show the downstream and upstream contributions, respectively. Because of  the  enhanced magnetic field, the synchrotron emission is higher  by a factor of $\approx 10$ downstream, as compared to upstream. At the same time, the synchrotron peaks of the SED both in downstream and upstream regions are located around 100 keV (slightly below and above 100 keV, respectively); the larger magnetic field  downstream is  effectively compensated by the shift of the maximum of the electron distribution upstream to higher energies (see   Fig. \ref{fig:cooledspt}\textbf{b}).
Due to the homogeneous distribution of  the target photons for the IC scattering (2.7 K CMBR), 
the resulting IC gamma-ray spectra essentially mimic the electron distributions. In particular, the 
Compton peak in the upstream region is noticeably shifted compared to the position of the 
Compton peak characterising the  downstream region. 

Note that unlike  previous calculations of proton spectra in galaxy clusters, we do not assume an \textit{a priori} spectrum, e.g. of the ``standard" form $E^{-2}exp[-E/E^*]$,  rather calculate its shape within the model of DSA, taking into account energy losses. Interestingly, our self--consistent calculation shows a less steep fall off in the final emitted radiation spectrum above the cut-off energy. The effect is the result of the combination of two factors: \textit{i}) the accurate treatment of DSA correctly taking into account energy losses, \textit{ii}) the accurate calculation of the shape of the produced electron spectrum. In this regard we note that  the commonly used delta-function approximation significantly deviates from the 
results of accurate calculations \citep{kelner}.

While some hints of  non-thermal hard X-ray emission have been reported from Coma and Ophiuchus clusters
(e.g. \citealp{ff,eckert}), these objects have not yet been detected in gamma-rays. Our calculations show that 
if protons are effectively accelerated by accretion shocks, one should expect both 
hard X-rays and TeV gamma-rays from nearby and powerful clusters. The detectability of the 
X-ray and gamma-ray  fluxes depend on the total energy  of protons accumulated in the cluster 
over the lifetime of 
the source (up to 10 Gyr)  and the  distance to the source. Also the angular size of the emission is likely to play a role, since nearby clusters of galaxies will appear as quite extended sources. Normalising the total energy of protons in the cluster to $10^{62}$ erg, the luminosities in both  synchrotron and IC components  of  emission are expected at the level of  $L\sim 10^{42}$ erg/s. To estimate the expected flux, $F_E=L/4\pi d^2$, one has to fix the distance to the object. For example, in the case of the Coma cluster ($d=100$ Mpc)  the results are  shown in Fig. \ref{fig:radcoppie}\textbf{b}, where the effect of gamma-ray photon absorption by the extragalactic background light (EBL)  has been taken into account. In the calculation we used the recent  EBL flux  from \citealp{franceschini}. The energy flux for the synchrotron component appears to be of the order of $10^{-12}$ erg cm$^{-2}$ s$^{-1}$, while the IC gamma-ray flux is approximately one order of magnitude less, given that gamma-rays with  energy exceeding 10 TeV are effectively absorbed in the EBL. The detection of these fluxes from extended regions such as clusters of galaxies is not an easy task, but 
possibly feasible with the next generation hard X-ray and TeV gamma-ray detectors.

\section{Shock Modification}
\label{Shock Modification}

%%%%%%%%%%%% fig.8
\begin{figure}[]
\centering
\includegraphics[scale=0.4]{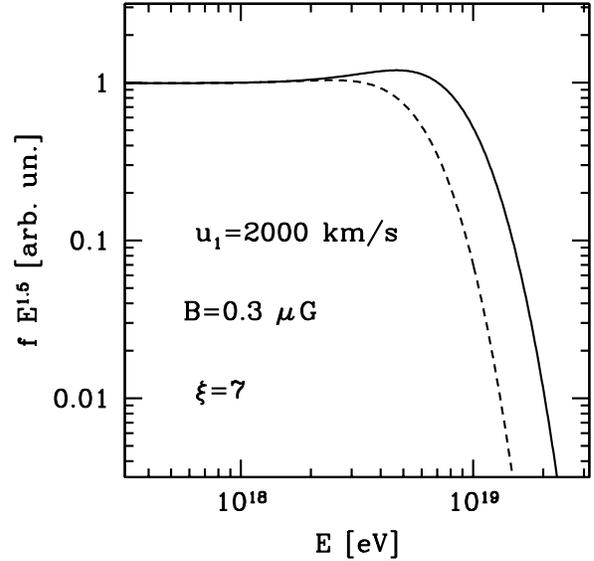}
\caption{\footnotesize{Proton spectra at the shock location for the case of a shock compression ratio $R=7$. The solid curve refers to an acceleration time of 10 Gyr, the dashed curve to an acceleration time of 5 Gyr. The shock velocity is 2000 km/s, the magnetic field upstream is $B_1=0.3$ $\mu$G and downstream $B_2=7 B_1 $.}}
\label{fig:shockPr7}
\end{figure}
%%%%%%%%%%%% fig.9
\begin{figure}[]
\centering
\includegraphics[scale=0.4]{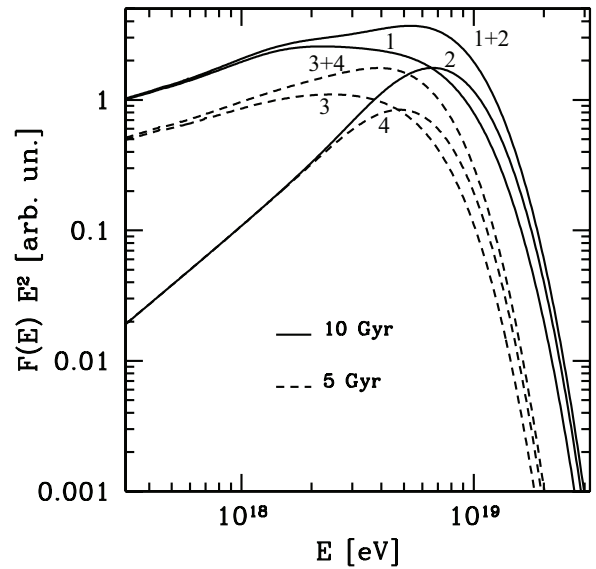}
\caption{\footnotesize{Spectra integrated over the upstream (2 and 4) and downstream (1 and 3) regions for the proton distributions in Fig. \ref{fig:shockPr7}. The solid lines correspond to an accelerator age of 10 Gyr and the dashed one to an age of 5 Gyr.}}
\label{fig:intPr7}
\end{figure}

We have shown in the previous sections that protons can be efficiently accelerated at cluster accretion shocks. In this scenario, the reaction of the accelerated particles on the structure of the shock itself can be quite significant, leading to
its modification. In turn, the modification of the shock reflects on the particle spectrum. This complex interplay leads to a variety of non-linear effects. A great deal of study has been dedicated to non-linear shock acceleration in recent years %\citep{d_v, kang, malkov1, ber_ell, blasi1} and 
(we refer to  \citealp{malk} for a review).
One of the consequences of shock modification is the increase of the acceleration efficiency coupled with a significant hardening  of the energy distribution of protons.  As a result, the  available kinetic energy transferred to accelerated particles is accumulated at the highest energies close to the cut-off. This part of the spectrum provides the dominant contribution to the pair production, and ultimately to the broad-band electromagnetic radiation emitted by the secondary electrons. Therefore the features of the shock modification should be reflected in the spectrum of the non-thermal electromagnetic radiation.

%%%%%%%%%%%% fig.10
\begin{figure}[]
\centering
\includegraphics[scale=0.4]{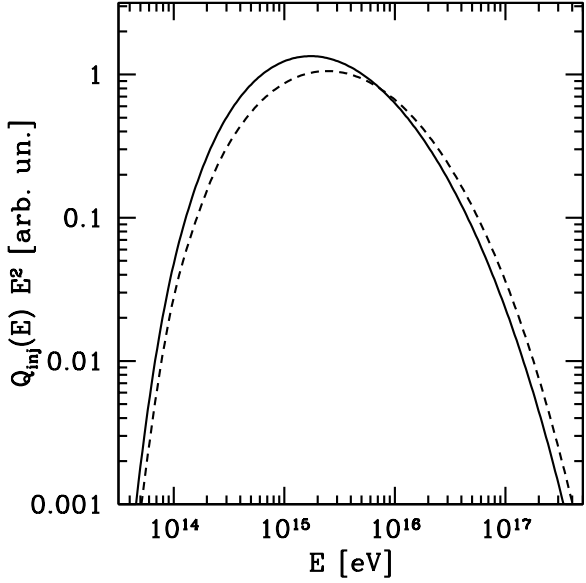}
\caption{\footnotesize{Pairs production spectra for an accelerator age of 5 Gyr. The solid line represents the downstream component, the dashed line refers to the upstream component.}}
\label{fig:sptcoppie7}
\end{figure}

%%%%%%%%%%%% fig.11
\begin{figure}[]
\centering
\includegraphics[scale=0.4]{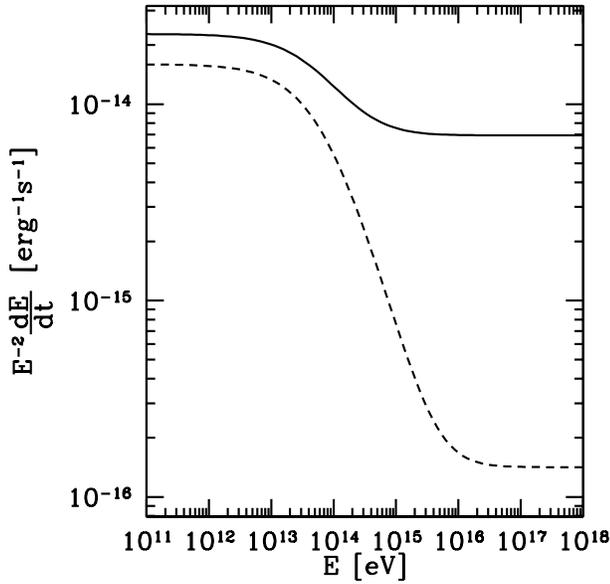}
\caption{\footnotesize{Energy loss rates for synchrotron plus IC losses. The magnetic field upstream is $B_1=0.3~\mu$G (dashed line), in the downstream region $B_2=7B_1$ (solid line).}}
\label{fig:losscoppie7}
\end{figure}

In order to explore this possibility, we take into account the modification of the shock in a simplified way, a full non-linear treatment being beyond the scope of the present study. We assume a compression ratio at the shock $R=7$, which results from the linear theory of DSA when the relativistic component of the fluid (i.e. the accelerated particles) dominate the system's dynamics. Analogously, the magnetic compression factor is increased to $\xi =$7. In this case the spectral slope predicted by the theory in the absence of losses is $\alpha = 1.5$. %(for a teview see \citet{malk}). 
The calculated proton spectrum at the shock surface, multiplied by $E^{1.5}$ is shown in Fig. \ref{fig:shockPr7} for a source age of 5 and 10 Gyr. One can see in the spectrum, just before the cutoff, a bump which is induced by the energy losses due to interactions with the CMBR.

In Fig. \ref{fig:intPr7} we show the spatially integrated spectra of protons in the upstream and downstream regions. The curves are multiplied by $E^{2}$ in order to emphasise the fact that the energy transferred to protons is accumulated just before the cut-off energy region. 

%%%%%%%%%%%% fig.12
\begin{figure}[]
\centering
\includegraphics[scale=0.4]{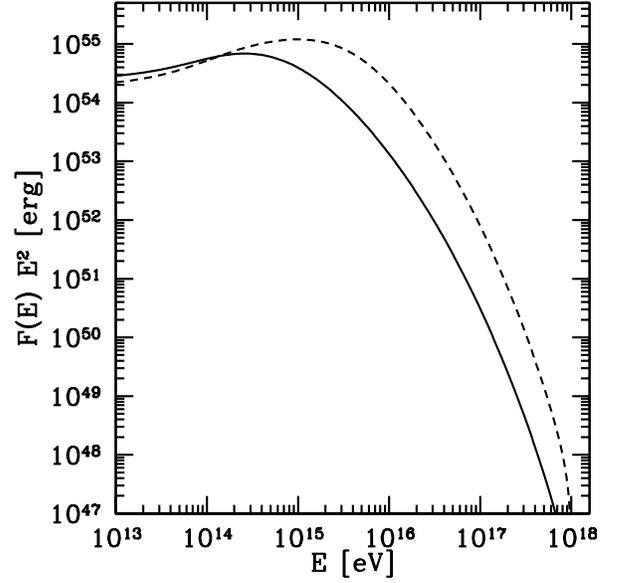}
\caption{\footnotesize{Electron spectra after cooling, downstream (solid line) and upstream (dashed line), obtained by considering the energy losses in Fig. \ref{fig:losscoppie7}.}}
\label{fig:cooledspt7}
\end{figure}
%%%%%%%%%%%% fig.13
\begin{figure}[]
\centering
\includegraphics[scale=0.5]{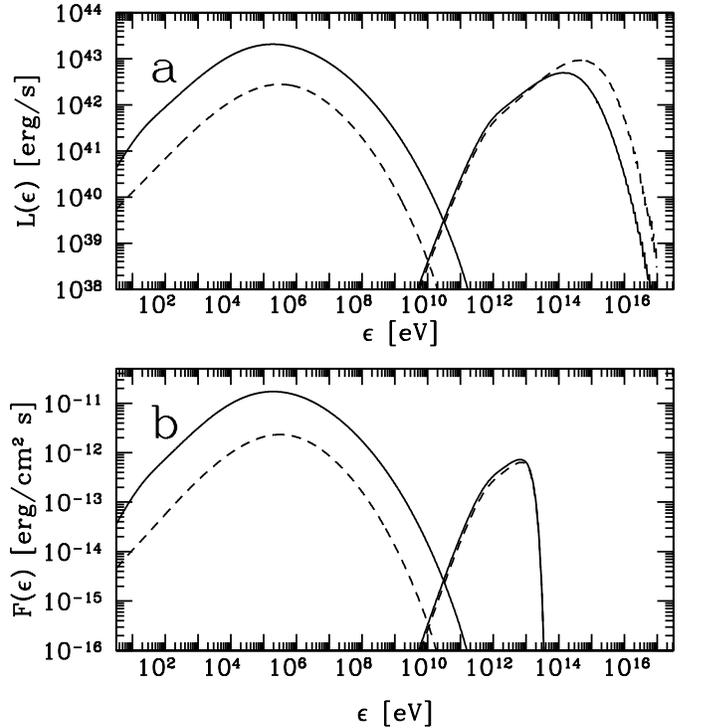}
\caption{\footnotesize{\textbf{a}) Broadband radiation spectra produced at the source by the electron distributions in Fig. \ref{fig:cooledspt7}\textbf{b}, downstream (solid line) and upstream (dashed line). \textbf{b}) Energy flux at the observer location, after absorption in the EBL, for a source distance of 100 Mpc.}}
\label{fig:radcoppie7}
\end{figure}

The secondary pair production spectra are shown in Fig. \ref{fig:sptcoppie7}, while Figure \ref{fig:losscoppie7} shows the energy loss rates for electrons upstream and downstream. Compared to Fig. \ref{fig:losscoppie}, the synchrotron cooling in the downstream region is increased  due to the higher magnetic field. This has an impact on the cooled spectra, as shown in Fig. \ref{fig:cooledspt7}. The spectra are normalised to a total energy released in high energy protons of $10^{62}$ erg.
The spectral characteristics are similar to those shown in Fig. \ref{fig:cooledspt}, but, due to the hard spectrum of protons, the amount of energy released in pairs is an order of magnitude larger than in the case of $R=4$.

This effect is directly reflected in the  broadband  radiation produced by secondary electrons 
(Fig. \ref{fig:radcoppie7}\textbf{a}).  The X-ray luminosity  is increased by an order of magnitude compared 
to the case of $R=4$.  The same applies for the IC radiation emitted in the upstream region.

Figure \ref{fig:radcoppie7}\textbf{b} shows the expected photon fluxes at Earth, for a source located at a distance $d=100$ Mpc. Because of the intergalactic absorption,  the gamma-ray flux above 10 TeV is dramatically suppressed. This leads  to the shift of the maximum of the arriving  IC gamma-ray spectrum  to  10 TeV 
with a flux at  the level of  $10^{-12} \ \rm  erg/cm^{2}s$.

\section{Conclusions}
\label{Conclusions}

Proton acceleration in galaxy clusters was studied in the framework of DSA via a detailed time-dependent numerical calculation that includes energy losses due to interactions of protons with photons of the CMBR. For realistic  shock speeds of a few thousand km/s and a background magnetic field close to  $1 \mu  \rm G$, the maximum energy achievable by protons is determined by the energy losses  due to  pair production and ranges from a few times $10^{18}$ eV to a few times $10^{19}$ eV.

We performed the calculations assuming that acceleration takes place on time scales comparable to the age of the cluster. Since steady state is never achieved in this scenario, a time-dependent treatment is required. Particle spectra, when calculated including the effect of energy losses, exhibit interesting features. The decay of the spectrum above the cut-off energy is not exponential. Its dependence on energy is shallower due to the flat profile of pair production timescales in the cut-off energy range. The time-dependent distributions of protons are used to calculate accurately the  production rates of secondary electron-positron pairs. These electrons cool rapidly via synchrotron radiation and IC scattering which proceeds in the Klein-Nishina  regime. The effect of the hardening induced by the KN cross-section is visible in the IC radiation spectra both in the upstream and downstream regions of the shock. For the fiducial Coma-like cluster used in this work, the synchrotron and IC peaks of the electron broadband SED  are at comparable levels and the associated flux from a source at the distance of a 100 Mpc is expected at the level of $10^{-12}$ erg cm$^{-2}$ s$^{-1}$ in the X-rays and an order of magnitude lower for TeV gamma-rays. Note that  although the maximum of the gamma-ray emission is located above 100 TeV, it unfortunately cannot be observed due to severe  intergalactic absorption. The expected  gamma-ray flux  from clusters of galaxies is at the limit of the sensitivity of present generation instruments, however it may be detectable with the future generation of detectors. The optimum energy interval for  gamma-ray detection  is between 1 and 10 TeV.

The  detectability of clusters in hard X-rays and  gamma-rays  associated with  interactions of ultrahigh energy protons with the CMBR, depends on the value of the parameter $A=W_{62}/d_{100}^2$, where $W_{62}=W/10^{62} \ \rm erg$ is the total energy released in cosmic rays normalised to $10^{62}$ erg, and $d_{100}=d/100$ Mpc is the distance normalised to 100 Mpc.
Obviously, the best candidates  for detection are nearby  rich galaxy clusters like Coma and Perseus located at distances $d \sim 100$ Mpc. On the other hand, due to the large extension of the  non-thermal  emission  (as large as several Mpc), and given that the angular size of the source $\theta \propto 1/d$,  the  probability of detection reduces  with the distance slower than  $1/d^2$. Nevertheless, as long as the total energy in accelerated protons  does not significantly exceed $10^{62}$ erg, the visibility of  clusters of galaxies in X-rays and gamma-rays is limited by objects located 
within a few 100 Mpc.

The chances of detection of non-thermal emission of clusters related to ultrahigh energy protons, especially in the hard X-ray band, can be significantly higher if protons are accelerated by  non-linear shocks modified by the pressure of relativistic particles.
In this scenario, a large fraction of the energy of the shock is transferred to relativistic protons. Moreover, in this case a very hard spectrum of protons is formed, thus the main fraction of non-thermal energy is carried by the highest energy particles.  These two factors can enhance the luminosity of X- and gamma-ray emission of secondary electrons by more than an order of magnitude, and thus  increase the probability of detection of clusters located beyond 100 Mpc.

In this paper we do not discuss the gamma-ray production related to interactions of accelerated protons with the ambient gas which can compete with the inverse Compton radiation of pair produced electrons. The relative contributions of these two channels depends on the density of the ambient gas and the spectral shape of accelerated protons.
The flux of gamma-rays from {\it pp} interactions can be easily  estimated based on the cooling time of protons, $t_{\rm pp} \approx 1.5 \times 10^{19}/n_{-4}$ s, where $n_{-4}=n/10^{-4}$ cm$^{-3}$ is the density of the ambient hydrogen gas, normalised to $10^{-4}$ cm$^{-3}$.  Then, the energy flux of gamma-rays at 1 TeV is estimated as $F_\gamma (\sim 1 \rm TeV) \approx  6 \times 10^{-12} \kappa W_{62} n_{-4}/d_{100}$ erg cm$^{-2}$ s$^{-1}$, where $\kappa$ is the fraction of the total energy of accelerated protons in the energy interval between 10 to 100 TeV  (these protons are primarily responsible for production of gamma-rays of energy $\sim 1$ TeV). For a proton energy spectrum extending to $10^{18}$ eV, this fraction is of order of $\kappa \sim 0.1$. Thus for an average gas density in a cluster like Coma, $n \sim 3 \times 10^{-4}$ cm$^{-3}$, the gamma-ray flux at 1 TeV is expected at the level of $10^{-12}$ erg cm$^{-2}$ s$^{-1}$ which is comparable to the contribution of IC radiation of secondary electrons. In the case of harder spectra of protons accelerated by non-linear shocks, the contribution of gamma-rays from {\it pp} interactions is dramatically reduced and the contribution of  secondary pairs to gamma-rays via IC scattering  strongly dominates over gamma-rays from {\it pp} interactions.

\begin{acknowledgements} 
We would like to thank O. Zacharopoulou and A. M. Taylor for fruitful discussions. GV acknowledges support from the International Max-Planck Research School (IMPRS) Heidelberg. SG acknowledges the support of the European Community under a Marie Curie Intra-European fellowship.
\end{acknowledgements}

{}

\end{document}